\documentclass[amsmath,amssymb,prl]{revtex4}

\begin{document}
\widetext
\title{Origin of spontaneous violation of the Lorentz symmetry:
Vortices in the cosmos
\footnotemark[1]
\footnotetext[1]{This work was done under the auspices of the U. S.
Department of Energy. The present address of the first author is: Department of Physics and Astronomy, Rutherford Building, University of Canterbury, Private Bag 4800, Christchurch 8020, New Zealand. }
}
\author{D. V. Ahluwalia\ \dag\  and \ T. Goldman \ddag}
\affiliation{%
\dag MP-9, MS H-846, Nuclear and Particle Physics Research Group\\
 Los Alamos National Laboratory,
Los Alamos, New Mexico 87545, USA}

\affiliation{%
\ddag T-5, MS B-283, Medium Energy Physics Theory Group\\
 Los Alamos National Laboratory,
Los Alamos, New Mexico 87545, USA}

\begin{abstract}
By carefully studying the $(1,0)\oplus(0,1)$ representation space for massive
particles we  point to the existence  of certain inherent tachyonic
dispersion relations: $E^2=\vec p^{\,2}-m^2\,$. We put forward an
interpretation that  exploits these
``negative mass squared'' solutions; rotational invariance is spontaneously
broken. Relevance of these results to the vortices in the cosmos is pointed
out.
\end{abstract}
\maketitle
\vspace{14pt}

\hrule
\vspace{7pt}
\centerline{{\sc Journal Reference:} Mod. Phys. Lett. A8:2623-2630,1993}
\vspace{7pt}
\hrule

\vspace{36pt}
\centerline{\sc Note added fourteen years later (on 04 March 2007)}

\vspace{7pt}
Just as ``negative energy solutions'' of Dirac equation are re-interpreted as
antiparticles, similarly the possibility exists for re-interpreting the tachyonic 
dispersion relations of all $(j,0)\oplus(0,j)$ representation spaces via 
spontaneous Lorentz symmetry breaking.  In \textit{ Mod. Phys. Lett. 
A8:2623-2630,1993} we exhibited this explicitly for the $j=1$ representation 
space. The interest in this old subject\cite{Kostelecky:1988zi} has grown 
markedly in recent years as is evident from numerous theoretical and 
phenomenological works on the subject\cite{Dvali:2007ks,Bertolami:2007mn,
Soda:2006wr,Chkareuli:2006yf,Ferreira:2006ga,Dubovsky:2006vk,
Cheng:2006us,Kostelecky:2005ic,Grossman:2005ej,Graesser:2005bg,
Arkani-Hamed:2004ar,Jenkins:2003hw,Moffat:2002nu,Chkareuli:2001xe,
Higashijima:2001sq,Kostelecky:2000mm,Colladay:1999jh,Bertolami:1998dn,
Hosotani:1994sc}. With this observation, we make this replacement of our
paper fourteen years after its initial publication. The Abstract and main text 
remain unaltered. The title is changed to reflect the underlying physics more 
closely.
\vspace{7pt}
\hrule
\newpage

The ubiquitous presence of  astrophysical vortices, such as seen in the
rotation of stellar objects and galaxies, is almost invariably thought of as
having its origin in the  turbulent hydrodynamical phenomenon  fueled initially
by gravity and later (as matter began to condense into large scale structures
in the universe)  by the interplay of  thermonuclear processes  and gravity.
Till recently, the earlier suggestions of Gamow \cite{GG} and G\"odel \cite{KG}
regarding a possible universal cosmic vorticity were not taken seriously
because it was thought \cite{SH} that such a vorticity is essentially ruled out
by the remarkable isotropy of the cosmic microwave background radiation (the
anisotropies \cite{CMBR} on all angular scales $\ge 7^\circ$ are $\Delta T/T
\approx 10^{-5}$). However, about a decade ago  Birch \cite{PB} noted that the
difference between the position angles of elongation and polarisation of high
luminosity classical-double radio sources are highly organised and interpreted
this observation as implying a universal cosmic vorticity $\sim\,10^{-13}\,
\mbox{rad yr}^{-1}\,$. Objections to Birch's work were immediately raised
\cite{OBJa,OBJb} (See Ref. \cite{reply} for Birch's reply to Ref. \cite{OBJb}).
But now Birch and his collaborators \cite{col} along with other independent
\cite{KY,A} researchers seem to have overcome the initial queries. Furthermore,
Obukhov \cite{YO} has very recently argued that it is not the cosmic vorticity
that leads to anisotropy in the cosmic microwave background radiation (and
other pathologies previously thought to be associated with cosmic vorticity),
but the non-vanishing of the shear tensor [13, Sec. 9.2] $\sigma_{ab}\,$.
Obukhov also provides a class of shear-free ($\sigma_{ab}=0$) homogeneous
models for cosmologies with cosmic vorticity, and via an analysis of the
polarisation rotation and red shift for certain radio sources reaches
conclusions very similar to Birch. Despite these remarks it is perhaps too
early to unambiguously interpret the positive \cite{PB,col,KY,A,YO}, or the
negative \cite{OBJb}, reports of cosmic vorticity, without further observations
and analysis.

Unlike the astrophysical vortices, it is difficult to interpret the universal
cosmic vorticity as having its origin in any ordinary turbulent hydrodynamical
process. Moreover, the astrophysical vortices themselves need not be the result
of hydrodynamical processes alone.
The study we present below may
 be a worthwhile exercise in its own
right,  providing us with a deeper understanding of the space time structure.
Following our recent work \cite{our} on the $(j,0)\oplus(0,j)$ representation
space, we will show by considering the $(1,0)\oplus(0,1)$ representation space
for the massive spin one particles in some detail that the Poincar\'e space
time symmetries may contain essential seeds for the violation of the rotational
symmetry. We realise that this is a somewhat surprising proposition, but this
result seems to follow so naturally from the general considerations below that
we believe that it may have some relevance for the observed vortices in the
universe.

In the notation of Ref. \cite{our},  fields in the  $(1,0)$ and $(0,1)$
representation space transform as
\begin{eqnarray}
&&(1,0):\quad \phi_R(\vec p\,)\
=\,\Lambda_R(\vec p\,)\,\phi_R(\vec 0\,) \,=\,\exp\left(+\,\vec
J\cdot\vec\varphi\,\right)\,\phi_R(\vec 0\,),\label{r}\\
&&(0,1):\quad\phi_L(\vec p\,)\,
=\,\Lambda_L(\vec p\,)
\,\phi_L(\vec 0\,) \,=\,\exp\left(-\,\vec J\cdot\vec\varphi\,
\right)\,\phi_L(\vec 0\,), \label{l}
\end{eqnarray}where $\vec J=3\times 3$ angular momentum matrices.
The boost parameter
$\vec \varphi$ is defined as
\begin{equation}
\cosh(\varphi)
\,=\,\gamma\,=\,\frac{1}{\sqrt{1-v^2}}\,=\,\frac{E}{m},\quad\quad \sinh(
\varphi)\,=\,v\gamma\,=\,\frac{| \vec p\, |}{m},\quad\quad\hat\varphi=
\frac{\vec  p}{| \vec p\,|},\label{bp}
\end{equation}
with $\vec p$ the three-momentum of the particle. In terms of the $(1,0)$
and $(0,1)$ fields given by Eqs. (\ref{r}) and (\ref{l}),
the chiral representation
$(1,0)\oplus(0,1)$ spinors are  defined as  six-element columns
\begin{equation}
\psi_{ch}(\vec p\,)\,=\,
\left(
\begin{array}{c}
\phi_R(\vec p\,)\\
\phi_L(\vec p\,)
\end{array}
\right).\label{chiral}
\end{equation}
To proceed further we introduce the generalised canonical representation
spinors
\begin{equation}
\psi_{ca}(\vec p\,)\,=\,
S\,\psi_{ch}(\vec p\,)\,=\,
\frac{1}{\sqrt 2}
\left(
\begin{array}{ccc}
\openone_3&{\;\;}&\openone_3\\
\openone_3&{\;\;}&-\openone_3
\end{array}
\right)\,\psi_{ch}(\vec p\,)
\,=\,
\left(
\begin{array}{cc}
\phi_R(\vec p\,)\,+\,\phi_L(\vec p\,)\\
\phi_R(\vec p\,)\,-\,\phi_L(\vec p\,)\\
\end{array}\right);\label{crs}
\end{equation}
with the notation $\openone_n=n\times n \;\mbox{identity matrix}$. Working in a
representation of the $\vec J$ matrices in which $J_z$ is diagonal, the rest
spinors $u_\sigma(\vec 0\,)$ and $v_\sigma(\vec 0\,)$, $\sigma=0,\pm 1 $, can
be written in the form the following six-element columns
\begin{subequations}
\begin{eqnarray}
&&u_{+1}(\vec 0\,)\,=\,
\left(
\begin{array}{c}
m\\
0\\
0\\
\{0\}
\end{array}
\right),\;
u_{0}(\vec 0\,)\,=\,
\left(
\begin{array}{c}
0\\
m\\
0\\
\{0\}
\end{array}\right),\;
u_{-1}(\vec 0\,)\,=\,
\left(
\begin{array}{c}
0\\
0\\
m\\
\{0\}
\end{array}\right)\quad,\label{rsu}\\
&&
v_{+1}(\vec 0\,)\,=\,
\left(
\begin{array}{c}
\{0\}\\
m\\
0\\
0
\end{array}
\right)\,,\,\,
v_{0}(\vec 0\,)\,=\,
\left(
\begin{array}{c}
\{0\}\\
0\\
m\\
0
\end{array}\right),\;
v_{-1}(\vec 0\,)\,=\,
\left(
\begin{array}{c}
\{0\}\\
0\\
0\\
m
\end{array}\right)\quad.\label{rsv}
\end{eqnarray}
\end{subequations}
In the above expression $\{0\}$ stands for a string of three vertical
``$0$.''Under the operation of parity $\cal P$:
$(1,0)\leftrightarrow (0,1)$.
Therefore, the $u_\sigma(\vec 0\,)$ spinors are {\it even} and
$v_\sigma(\vec 0\,)$ spinors are {\it odd} under the operation $\cal P$. This
even/odd-ness we call intrinsic spinor parity
\footnotemark[1]
 \footnotetext[1]{Intrinsic spinor parity is to be distinguished from the
relative intrinsic parity.
The reader may wish to refer to Ref. \cite{AJG} for details. }.
The explicit expressions for the $(1,0)\oplus(0,1)$ canonical representation
spinors $\psi_{ca}(\vec p\,)$ are obtained from the rest spinors, Eqs.
 (\ref{rsu}) and (\ref{rsv}),
by the action of the boost matrix $M(\vec p\,)$ obtained from the
transformation
properties (\ref{r}) and (\ref{l}) and the definition (\ref{crs}):
\begin{equation}
\left\{
\begin{array}{c}
u_\sigma(\vec p\,)\\
v_\sigma(\vec p\,)
\end{array}
\right\}
\,=\,
M(\vec p\,)
\,
\left\{
\begin{array}{c}
u_\sigma(\vec 0\,)\\
v_\sigma(\vec 0\,)
\end{array}
\right\}
\,=\,
\left(
\begin{array}{ccc}
\cosh(\vec J\cdot\vec\varphi)&{\,\,} &\sinh(\vec J\cdot\vec\varphi) \\
\sinh(\vec J\cdot\vec\varphi)&{\,\,} &\cosh(\vec J\cdot\vec\varphi)
\end{array}\right)\,
\left\{
\begin{array}{c}
u_\sigma(\vec 0\,)\\
v_\sigma(\vec 0\,)
\end{array}
\right\}.
\end{equation}
For the sake of completeness we parenthetically note the following
identities
\begin{subequations}
\begin{eqnarray}
&&\overline{u}_\sigma(\vec p\,)\,u_\sigma(\vec p\,) \,=\,+\,m^2\,,\quad
\overline{v}_\sigma(\vec p\,)\,v_\sigma(\vec p\,) \,=\,-\,m^2\quad,\\
&&P_u\,+\,P_v\,=\,\openone_6\,,\quad P_u^2\,=\,P_u
\,,\quad P_v^2\,=\,P_v\,,\quad P_u\,P_v\,=\,0\quad,
\end{eqnarray}
where
\begin{equation}
P_u\equiv +\, \frac{1}{m^2}\sum_{\sigma=0,\pm1} u_\sigma(\vec p\,)\,
\overline{u}_\sigma(\vec p\,)\,,\quad
P_v\equiv -\,\frac{1}{ m^2}\sum_{\sigma=0,\pm1} v_\sigma(\vec p\,)\,
\overline{v}_\sigma(\vec p\,)\quad;
\end{equation}
and we define
\begin{equation}
\overline{\psi}_\sigma(\vec p\,)\,\equiv\,\psi^\dagger_\sigma(\vec p\,)\,
\gamma_{00}\,,\quad
\gamma_{00}\,\equiv\,
\left(
\begin{array}{ccc}
\openone_3 &{\;\;} &0 \\
0 &{\;\;} &-\openone_3
\end{array}
\right)\quad.
\end{equation}
\end{subequations}

The hint for a  possible mechanism to  violate rotational invariance within the
context of space time symmetries now emerges by carefully examining the wave
equation satisfied by the $(1,0)\oplus(0,1)$ spinors. To derive the wave
equation we note that the general structure of the rest spinors given by Eqs.
(\ref{rsu}) and (\ref{rsv})  implies that
\begin{equation}
\phi_R(\vec 0\,)\, \,=\,\wp_{u,v}\,\phi_L(\vec 0\,).\label{puv}
\end{equation}
where
\begin{eqnarray}
\wp_{u,v}\,=\,\left\{
\begin{array}{l}
+\,1\;\;\mbox{for}\;\mbox{$u$-spinors}\;,\\
-\,1\;\;\mbox{for}\;\mbox{$v$-spinors}\quad.
\end{array}\right.\label{pm}
\end{eqnarray}
In a similar context for spin one half, Ryder [6, p.44] assumes the validity of
Eq. (\ref{puv}),
without the factor $\wp_{u,v}$,
on the grounds that ``when a particle is at rest, one cannot
define its spin as either left- or right-handed.'' However, as we note this is
simply a consequence of the general structure of our theory --- moreover, in
the process we find the additional  $\wp_{u,v}$ factor on the
{\it rhs} of Eq. (\ref{puv}). This factor has been found to have profound
significance
\footnotemark[2]
\footnotetext[2]{For  detailed  consequences of the factor $\wp_{u,v}$ the
reader is referred to Ref. \cite{AJG}.}
for the internal consistency and consequences
for the $(j,0)\oplus(0,j)$ quantum field theory.
Using Eqs. (\ref{r}), (\ref{l}), (\ref{puv}) and $\wp_{u,v}^2 =1$,
it can be shown that
\begin{subequations}
\begin{eqnarray}
&&\phi_R(\vec p\,)\,=\,\wp_{u,v}\,\exp\left(+\,2\,\vec
J\cdot\vec\varphi\,\right)
\phi_L(\vec p\,)\label{u}\\
&&\phi_L(\vec p\,)\,=\,\wp_{u,v}\,\exp\left(-\,2\,\vec
J\cdot\vec\varphi\,\right) \phi_R(\vec p\,).\label{d}
\end{eqnarray}
\end{subequations}
We now
\begin{enumerate}
\item[1.)] Expand the exponentials in
Eqs. (\ref{u}) and (\ref{d}) in terms of $\cosh\left( 2\,\vec
J\cdot\vec\varphi\,\right)$ and $\sinh\left( 2\,\vec
J\cdot\vec\varphi\,\right)$ and use the $j=1$ expansions
\begin{subequations}
\begin{eqnarray}
\cosh(2\vec J\cdot\vec\varphi)&\,=\,&1+2(\vec
J\cdot\hat p\,)(\vec J\cdot\hat
p\,)\sinh^2\varphi, \label{cpbe}\\
\sinh(2\vec J\cdot\vec \varphi)&\,=\,& 2(\vec J\cdot\hat
p\,)\cosh\varphi\sinh\varphi,
\label{cpbf}
\end{eqnarray}
\end{subequations}
along with the definition of  boost parameter $\vec \varphi$  given by Eq.
(\ref{bp});

\item[2.)] Multiply both sides of
Eqs. (\ref{u}) and (\ref{d})
by $m^{2}$ and use $\wp_{u,v}^2=1$;

\item[3.)] Note the definition of $\psi_{ch}(\vec p\,)$ Eq. (\ref{chiral}),
 or equivalently
observe that:
$
\psi_{ch}(\vec p\,)\,=\,S^{-1}\,\psi_{ca}(\vec p\,)\,.
$
\end{enumerate}
The enumerated steps after appropriate rearrangement
allow us to recast  Eqs. (\ref{u}) and (\ref{d})
into the following wave equation
\begin{equation}
\left(\gamma_{\mu\nu}\,p^\mu p^\nu\,-\,\wp_{u,v}\,m^2\,\openone_6
\right)\,\psi(\vec p\,)\,=\,0\label{eq}
\quad,
\end{equation}
with (chiral representation expression)
\begin{equation}
\gamma_{\mu\nu}p^\mu p^\nu\,=\,
\left(
\begin{array}{cc}
0&B\,+\,2\,(\vec J\cdot\vec p\,)\,p^0\\
B\,-\,2\,(\vec J\cdot\vec p\,)\,p^0 & 0
\end{array}
\right),\label{gma}
\end{equation}
where $B\,=\,\eta_{\mu\nu}\,p^\mu p^\nu \,+\, 2\,(\vec J\cdot\vec p\,)
\,(\vec J\cdot\vec p\,)\,$.
In  Eq. (\ref{eq})  we
have dropped the representation-identifying subscript ``ch'' because Eq.
(\ref{eq}) is valid for {\it all} spinors $\psi(\vec p\,)$ related via the
unitary transformation:
\begin{equation}
 \psi(\vec
p\,)\,=\,A\,\psi_{ch}(\vec p\,)\,,\quad
\gamma_{\mu\nu}\,=\, A\,
\left(\gamma_{\mu\nu}\right)_{ch}\,A^{-1}.
\end{equation}
The  wave equation (\ref{eq}), except for the already indicated factor of
${\wp_{u,v}}$ attached to the mass term, is identical to the Weinberg equation
\cite{SW}.

Now a surprising feature of the space time symmetries encompassed in Eq.
(\ref{eq}) emerges. According to Eq.
(\ref{eq}) the solutions $\psi(\vec p\,)$ can not only be interpreted (as
by construction) as associated with the Einsteinian dispersion relation
$E^2=\vec p^{\,2}+m^2$ but also with the
tachyonic dispersion relation $E^2=\vec p^{\,2}- m^2$. This is readily
inferred by studying the dispersion relations associated with Eq.
(\ref{eq}):
\begin{equation}
\mbox{Det}
\left(\gamma_{\mu\nu}\,p^\mu p^\nu\,-\,\wp_{u,v}\,m^2\,\openone_6
\right)\,=\,0.\label{det}
\end{equation}
The {\it lhs} of Eq. (\ref{det}) is a ${12^{\mbox{th}}}$ order polynomial
in $E$, and can be factorised to yield the dispersion relations:
\begin{equation}
E^2\,=\,\vec p^{\,2}\,-\,\wp_{u,v}\,m^2\,,\quad
E^2\,=\,\vec p^{\,2}\,+\,\wp_{u,v}\,m^2,\label{dr}
\end{equation}
each with a multiplicity three.

Parenthetically, it may be noted that the set of dispersion relations
(\ref{dr})
is invariant under $\wp_{u,v} \rightarrow -\wp_{u,v}\,$. This invariance arises
from the fact that ${\mbox{Det}} \left(\gamma_{\mu\nu}\,p^\mu
p^\nu\,-\,\wp_{u,v}\,m^2\,\openone_6 \right) $ depends only on the {\it even }
powers of $\wp_{u,v}$.

It  is surprising
that the space time symmetries associated with the massive $(1,0)
\oplus(0,1)$ representation space includes in it the tachyonic dispersion
relations, $E^2=\vec p^{\,2}-m^2$. However, this inclusion
 may provide the sought after
physical origin  [18, Sec. 1] of the  tachyonic mass (the ``{\it negative}
mass squared'' of Ref. \cite{JB}) when one adds
the quartic self-coupling which appears
in the simplest versions of field theories with broken symmetry.  Therefore,
let us postulate  the following Lagrangian density for the self-coupling in the
 $(1,0)\oplus(0,1)$ representation space:
\begin{equation}
{\cal L}(x)\,=\,
-\,{\wp_{u,v}}\,\overline{\psi}(x)\,\gamma_{\mu\nu}\,\partial^\mu\partial^\nu\,
\psi(x) - \left\{m^2\,\overline{\psi}(x)\psi(x)\,+\,\lambda\,
\left(\overline{\psi}(x)\psi(x)\right)^2\right\}.\label{ld}
\end{equation}
In the absence of the self-coupling (i.e. for $\lambda=0$) the above Lagrangian
density yields Eq. (\ref{eq})
\footnotemark[3]
\footnotetext[3]{
To see this the reader should note: $\wp_{u,v}^2=1$ and $\psi(x) \equiv
\exp(-i\wp_{u,v}\, p\cdot x)\,\psi(\vec p\,)\,$. Here $\psi(\vec p)$ stands
either for $u_\sigma(\vec p\,)$ or $u_\sigma(\vec p\,)$  --- the sign in the
exponential $\exp(-i\wp_{u,v}\, p\cdot x)$ is chosen
accordingly. }.

The solutions with $m^2 > 0\,$, implicit in the Einsteinian relation $E^2=\vec
p^{\,2} + m^2\,$, corresponding to the minimum of the field-space potential
\begin{equation}
V(\psi,\overline{\psi})\equiv
m^2\,\overline{\psi}(x)\psi(x)\,+\,\lambda\,
\left(\overline{\psi}(x)\psi(x)\right)^2,
\label{fsp}
\end{equation}
are [now interpreting $\psi(x)$ to be a field operator: $\Psi(x)$]  the
$(1,0)\oplus(0,1)$ particles ${\cal N}^\pm$  of mass $m$ associated with
$\langle{\mbox{vac}}|\overline{\Psi}(x)\Psi(x)|{\mbox{vac}}\rangle=0\,$. For
the solutions with $m^2 < 0$, the minimum of $V(\psi,\,\overline\psi)$ lies
at
\begin{equation} 
\overline{\psi}(x)\,\psi(x) \,=\,-\,\frac{m^2}{2\,\lambda}
\,=\,\,\frac{|m^2|}{2\,\lambda}.\label{min}
\end{equation}
Quantum field theoretically, when $\psi(x)$ is  considered  as a field operator
$\Psi(x)$, we interpret Eq. (\ref{min}) as: $\langle\mbox{vac}\vert{\overline
\Psi}(x)\Psi(x) \vert\mbox{vac}\rangle = |m^2|/{2\lambda}\,$. But, there are
two possibilities for the vacuum expectation value of the field operator
\begin{quote}
\begin{enumerate}
\item[Case I:]
$\langle\mbox{vac}\vert\Psi(x) \vert\mbox{vac}\rangle\,=\,0 \,;$ or
\item[Case II:]
$\langle\mbox{vac}\vert\Psi(x) \vert\mbox{vac}\rangle\,=\,\sqrt{|m^2|/2\lambda}
 \,.$
\end{enumerate}
\end{quote}
Case I has already been considered elsewhere \cite{AGJ}.
Here,  we confine our attention to Case II.

The degenerate vacua (obtained by a ${\overline \psi}(x)\psi(x)$ preserving
rotation in the field space)  are related by a continuous {\it global} gauge
symmetry of the Lagrangian density (\ref{ld}) under $\psi(x) \rightarrow
e^{i\,\Lambda}\,\psi(x)$, $\Lambda=$ constant.
A {\it particular} choice of the vacuum involves a
particular convention for the values of the field, say
\footnotemark[4]
\footnotetext[4]{Note that in Eq. (\ref{mm}) the first term on
the {\it rhs} is to be interpreted as a six-column element with all
entries equal to zero except one, which equals $\sqrt{|m^2|/2\lambda}\,$.}
\begin{subequations}
\begin{eqnarray}
\psi(x)&\,=\,&
\sqrt{\frac{|m^2|}{2\,\lambda}}\,+\,
{\frac{\psi^{M_+}(x)\,+\,\psi^{M'_+}(x) }{\sqrt{2}}},\label{mm}\\
&\,\equiv\,&
\rho(x)\,\exp\left[i\theta(x)\right],\label{mmm}
\end{eqnarray}
\end{subequations}
with $\langle{\mbox{vac}}|\Psi^{M_+}(x)|{\mbox{vac}}\rangle=0=
\langle{\mbox{vac}}|\Psi^{M'_+}(x)|{\mbox{vac}}\rangle\,$. Here $\psi^{M_+}(x)$
and $\psi^{M'_+}(x)$ are the Majorana-like spinors \cite{AGJ} in the
$(1,0)\oplus(0,1)$ representation space. In writing (\ref{mm}) we have further
exploited the freedom to choose a representation in which the Majorana-like
spinors are {\it real}.

The choice (\ref{mm}) obviously breaks the original symmetry of the Lagrangian
density, and chooses ``at random'' a particular direction in the
$(1,0)\oplus(0,1)$ representation space. This ``particular direction in the
$(1,0)\oplus(0,1)$ representation space,'' unlike its counter part in the
complex $\phi^4$-theory with quartic coupling,  breaks the
rotational symmetry.

We now follow the standard logic \cite{TG}
to conclude that the $m^2<0$ solutions can now be interpreted as:
\begin{enumerate}
\item[1.)] A (Goldstone-like) spin one massless  Majorana-like particle
$\hat\eta$.
Physically this can be seen as a quanta of ``angular oscillations'' at the
minimum of potential $V(\psi,{\overline \psi})$,  Eq. (\ref{fsp}),
corresponding
to $\overline{\psi}(x)\,\psi(x)$ preserving rotations in the field space.

\item[2.)] A massive Majorana-like spin one particle ${\cal N}^0$
 of mass ${\sqrt 2}
|m|$.
Physically this can be seen as a quanta of ``radial oscillations'' in
$V(\psi,{\overline \psi})$, Eq. (\ref{fsp}), arising from
$\overline{\psi}(x)\,\psi(x)$-non-preserving fluctuations about the
minimum of
$\overline{\psi}(x)\,\psi(x)$ given by Eq. (\ref{min}).
\end{enumerate}

No known set of particles can be associated with the quartet of particles: $\{
{\cal N}^\pm,\;{\cal N}^0,\;\hat\eta\}\,$. A priori, it is also difficult to
put any constraints on the mass parameter $m$ associated with the massive
sector of this quartet. Nevertheless,  a possible cosmic vorticity seems to be
a natural candidate to be associated with the built-in breaking of the
rotational symmetry in our interpretation of the $(1,0)\oplus(0,1)$
representation space. While it may be interesting to speculate on the possible
existence (and its consequences for the stellar and galactic formation) of
space time domains with broken rotational symmetry (each domain, like in a
ferromagnet, characterised by a unique direction) we find these matters beyond
the scope of the present paper. The main result, and somewhat of a surprise, is
the interpretational structure which seems to be naturally associated with the
$(1,0)\oplus(0,1)$ representation space. This interpretation is essentially
forced upon us by the inherent existence of the tachyonic dispersion relation:
$E^2=\vec p^{\,2} -m^2\,$. It may not be irrelevant to recall that when a
similar study  \cite{PAMD}
was carried out for the  $(1/2,0)\oplus(0,1/2)$ representation
space ``negative energy solutions'' are forced upon us --- the
reinterpretation of these unexpected solutions lead to the prediction and
discovery of ``anti-particles.'' The study of the $(1,0)\oplus(0,1)$
representation space seems to contain in it the sought after physical origin of
the the ``{\it negative} mass squared'' which appears in field theories with
broken symmetries with quartic self-coupling  --- it can be shown that this
negative mass squared (or tachyonic dispersion relation) is a common feature of
all $j\ge 1$ {\it bosonic} $(j,0)\oplus(0,j)$ representation spaces.

In conclusion we have noted
that  the free, i.e. without any interactions,
$(1,0)\oplus(0,1)$ representation for massive particles contains certain
inherent tachyonic elements. The ``negative mass squared,'' which emerges
is reminiscent of $m^2<0$ of the
simplest versions of quantum field theories with
spontaneous symmetry breaking. This observation led us to introduce a quartic
self-coupling in the $(1,0)\oplus(0,1)$ representation space. We showed that
 this
 representation space with a quartic self-coupling may be
interpreted as a quartet of particles $\{ {\cal N}^\pm,\;{\cal
N}^0,\;\hat\eta\}\,$, of  masses $m_{{\cal N}^\pm}=m,\; m_{{\cal N}^
0}=\sqrt{2}\, m,\; m_{\hat\eta}=0,\,$ but with a spontaneously broken
rotational
symmetry. However,
the observation of Birch \cite{PB} and others \cite{col,KY,A,YO} on
cosmic vorticity, and the empirical existence of stellar and galactic vortices,
may be possible manifestations of this underlying breaking of rotational
symmetry in the $(1,0)\oplus(0,1)$ representation space  with quartic coupling.

\section*{Acknowledgements}
We wish to register our {\it zimpoic} thanks to Mikkel Johnson for asking the
right questions at the right time.

\end{document}